\documentclass[fleqn,usenatbib]{mnras}

\usepackage{newtxtext,newtxmath}

\usepackage[T1]{fontenc}
\usepackage{ae,aecompl}
\usepackage{graphicx, subfigure}
\usepackage{float}

\usepackage{graphicx}	
\usepackage{amsmath}	
\usepackage{amssymb}	

\title[Likelihood of Detecting Young Giant Planets]{The Likelihood of Detecting Young Giant Planets with High Contrast Imaging and Interferometry}

\author[A.L. Wallace et al.]{A.L. Wallace$^{1}$\thanks{E-mail: alexander.wallace@anu.edu.au}, M.J. Ireland$^{1}$
\\
$^{1}$Research School of Astronomy \& Astrophysics, Australian National University, Canberra, ACT 2611, Australia}

\pubyear{2018}

\begin{document}
\label{firstpage}
\pagerange{\pageref{firstpage}--\pageref{lastpage}}
\maketitle

\begin{abstract}
Giant planets are expected to form at orbital radii that are relatively large compared to transit and radial velocity detections (>1\,AU).  As a result, giant planet formation is best observed through direct imaging.  By simulating the formation of giant (0.3--5$M_{J}$) planets by core accretion, we predict planet magnitude in the near infrared (2--4 $\mu$m) and demonstrate that, once a planet reaches the runaway accretion phase, it is self-luminous and is bright enough to be detected in near infrared wavelengths. Using planet distribution models consistent with existing radial velocity and imaging constraints, we simulate a large sample of systems with the same stellar and disc properties to determine how many planets can be detected.  We find that current large (8--10m) telescopes have, at most a 0.2\% chance of detecting a core accretion giant planet in the L' band and 2\% in the K band for a typical solar type star.  Future instruments such as METIS and VIKiNG have higher sensitivity and are expected to detect exoplanets at a maximum rate of 2\% and 8\% respectively.
\end{abstract}




\section{Introduction}
Giant planets are thought to form primarily by the core accretion model at $\sim$3--30\,AU separations from their stars \citep{alibert2005models}.  To properly understand planet formation, it is necessary to observe exoplanets as they form.  Due to the relatively large orbital radii involved (>5\,AU) transit and radial velocity methods are inadequate and these planets must be observed through direct imaging.  With some notable exceptions (\citet{marois2008direct}, \citet{lagrange2009probable}, \citet{macintosh2015discovery}) there has been little success in discovering planets by this method, mostly due to limitations in sensitivity and angular resolution.  However, the recent discovery of PDS 70 b \citep{keppler2018discovery} arguably provided the first clear image of a planet in the process of forming.

Previous direct imaging surveys, despite failing to detect planets, have provided insight into the detection limits of current instruments and techniques.  Using models of planet luminosity, some have provided limits on detectable planet mass.  Surveys such as \citet{lafreniere2007gemini}, \citet{nielsen2008constraints} and \citet{biller2013gemini} were able to establish detection limits of 2, 4 and 1M$_{\mathrm{J}}$ respectively for the most ideal targets in their samples.

The mass limits from these surveys, combined with the few detections of high mass planets can provide information about the mass/orbital radius distribution.  While radial velocity surveys such as \citet{cumming2008keck} can constrain the distribution of small separation planets, direct imaging surveys such as those presented in \citet{brandt2014statistical} can be applied to wide separations.  Both of these surveys provided functional forms for planet frequency as a function of mass and orbital radius, however, the limited direct imaging data only provides weak constraints for wide separations.  Most direct imaging surveys provide an estimate of how common planets are at wide separations. 

The two papers with the most stringent occurrence rates that overlaps with core-accretion separation ranges listed in the review of \citet{bowler2018occurrence} are \citet{brandt2014statistical} and \citet{vigan2017vlt}. \citet{brandt2014statistical} concluded that 1.0--3.1\% of stars have a high mass (>5\,$M_{J}$) planet from 10--100\,AU, while concluded that 0.25--5.55\% of stars (95\% confidence) have a high mass (5--14\,$M_{J}$) planet from 5--500\,AU (with significant extrapolation to separations below 20\,AU).  More recently, \citet{nielsen2019gemini} found an occurence rate of 9\% for high mass planets (5--13\,$M_{J}$) at 10-100\,AU around high mass stars (>1.5$M_{\odot}$) observed by the Gemini Planet Imager Exoplanet Survey.

Whether a planet can be detected is determined by the planet's luminosity and the sensitivity of the instrument.  A planet's luminosity depends on its mass and internal entropy \citep{mordasini2017characterization}, which is in turn determined by its age and also the process by which it was formed \citep{marley2007luminosity,szulagyi2016thermodynamics}.  A young planet emits most of its formation energy in the infrared \citep{fortney2008synthetic,eisner2015spectral}.  Planet formation models have giant planets forming in a protoplanetary disc either by core accretion (timescale of order Myr) or disc instability (timescale of order kyr.)  Planets formed by core accretion have evolution bounded by low entropy (cold start) models or high entropy (hot start) models.  If the accretion shock is radiatively efficient, the planet loses most of its heat and evolves according to cold start models \citep{baruteau2016formation}.  If the shock is radiatively inefficient, the planet retains the heat and evolves according to hot start models leading to a higher initial luminosity \citep{mordasini2012characterization}.  This tends to happen if the planetary core is sufficiently massive \citep{mordasini2013luminosity}.

Planets formed by disc instability have no accretion shock and are generally expected to retain their heat and thus evolve through hot start models unless the protoplanet continues to accrete gas after collapse \citep{baruteau2016formation}.  Systems of bright, wide companions such as HR 8799 are often thought to have been formed by disc instability \citep{boss2011formation} but these systems are also quite rare \citep{forgan2013towards} and core accretion is the favoured formation model for the gas giants in our own solar system \citep{d2010giant}.  The luminosity of giant planets has been studied previously with a range of extreme assumptions such as the hot start \citep{baraffe2003evolutionary} and cold start models \citep{marley2007luminosity}, as well as a combination of the two (warm start, \citet{spiegel2012spectral}.)

In summary, previous work has used observations to constrain the distribution of planets while other work has attempted to predict planet magnitude during and after formation.  However, this is the first time these methods have been combined with observational constraints in order to predict exoplanet detection yield.  In this paper, we develop a plausible framework for determining the detectability of giant planets from existing and planned instruments.  We simulate planet formation by core accretion based on models from \citet{lissauer2009models} using appropriate parameters of protoplanetary discs based on \citet{tripathi2017millimeter}.  We determine the brightness of the planet based on \citet{zhu2015accreting} and \citet{spiegel2012spectral}.  Combining this with an assumed distribution of existing planets, we create a realistic sample of young core-accretion giant planets around Sun-like stars.  Using observed contrast limits from existing instruments such as NIRC2 and NaCo and theoretical limits for planned projects such as Hi-5/VIKiNG with VLTI \citep{defrere2018path} and METIS with the ELT, we calculate the probability of planet detection by direct imaging.  Due to the fact that, to our knowledge, this is the first time observational constraints have been added, we are deliberately simplistic with the theoretical aspects of planet formation and could extend this model in future work.

\section{Distribution of Existing Planets}
\label{sec:dist}
Due to the relatively few detections of wide separation planets (e.g. $M<5M_J$, 20\,AU$<r_{\rm{orb}}<$100\,AU), their distribution is difficult to constrain.  However, there have been studies of radial velocity data to determine the distribution of short period planets around sun-like stars  \citep{cumming2008keck} and higher mass stars \citep{bowler2009retired}.  More recently, \citet{mulders2018exoplanet} developed a population simulator based on the latest Kepler data release and  \citet{fernandes2019hints} derived a broken power law distribution with a turnover at the snow line based on radial velocity observations from \citet{mayor2011harps}.

The functional form of the distribution is given by
\begin{equation}
\frac{dN}{d\mathrm{ln}Md\mathrm{ln}P}=C'M^{\alpha}P^{\beta'}
\end{equation}
where $N$ is the number of planets per system and $C'$ is a normalisation constant.

This can be converted to be in terms of  orbital radius giving
\begin{equation}
\frac{dN}{d\mathrm{ln}Md\mathrm{ln}r_{orb}}=CM^{\alpha}r_{orb}^{\beta}
\label{eq:dist1}
\end{equation}
where $\beta=3\beta'/2$ due to conversion between period and orbital radius.  In the broken power law distribution, $\beta$ is expected to be constant at $\beta_{1}$ for a set of small orbital radii and suddenly change at a given turnover point before remaining constant at $\beta_{2}$ for larger radii.  In \citet{fernandes2019hints}, two broken power laws are derived: one is the symmetric broken power law where $\beta_{2}=-\beta_{1}$ and the slightly more accurate asymmetric law where $\beta_{1}$ and $\beta_{2}$ are independent of each other.  In this paper, we test out both of these distributions using $\alpha=-0.45$ \citep{fernandes2019hints}.  For the symmetric case, $\beta_{1} = 0.945$ and $\beta_{2}=-\beta{1}$ with a turnover at 1.77\,AU.  The asymmetric case has $\beta_{1}=0.795$ and $\beta_{2}=-1.83$ with a turnover at 2.8\,AU.

In this simulation we assume a planet mass range of 0.3--5\,M$_{\rm{J}}$.  This mass range is slightly arbitrary.  Detectability also depends on whether the protoplanetary disc is optically thin.  If there is substantial grain growth, by the time giant planets form, the disc is optically thin.  On the other hand, if grain growth is neglibible, even with gas accretion, the disc is optically thick in the line of sight of the planet.  This scenario is adopted in \citet{szulagyi2017observability} and \citet{szulagyi2019observability}.  We assume an intermediate scenario in which planets greater than 0.3\,M$_{\rm{J}}$ are considered massive enough to produce a gap which is optically thin so are detectable during formation.  Above 5\,M$_{\rm{J}}$, planets are too rare to be worth simulating for a small to moderate sample size.

\begin{figure}
\centering
\includegraphics[width=8cm]{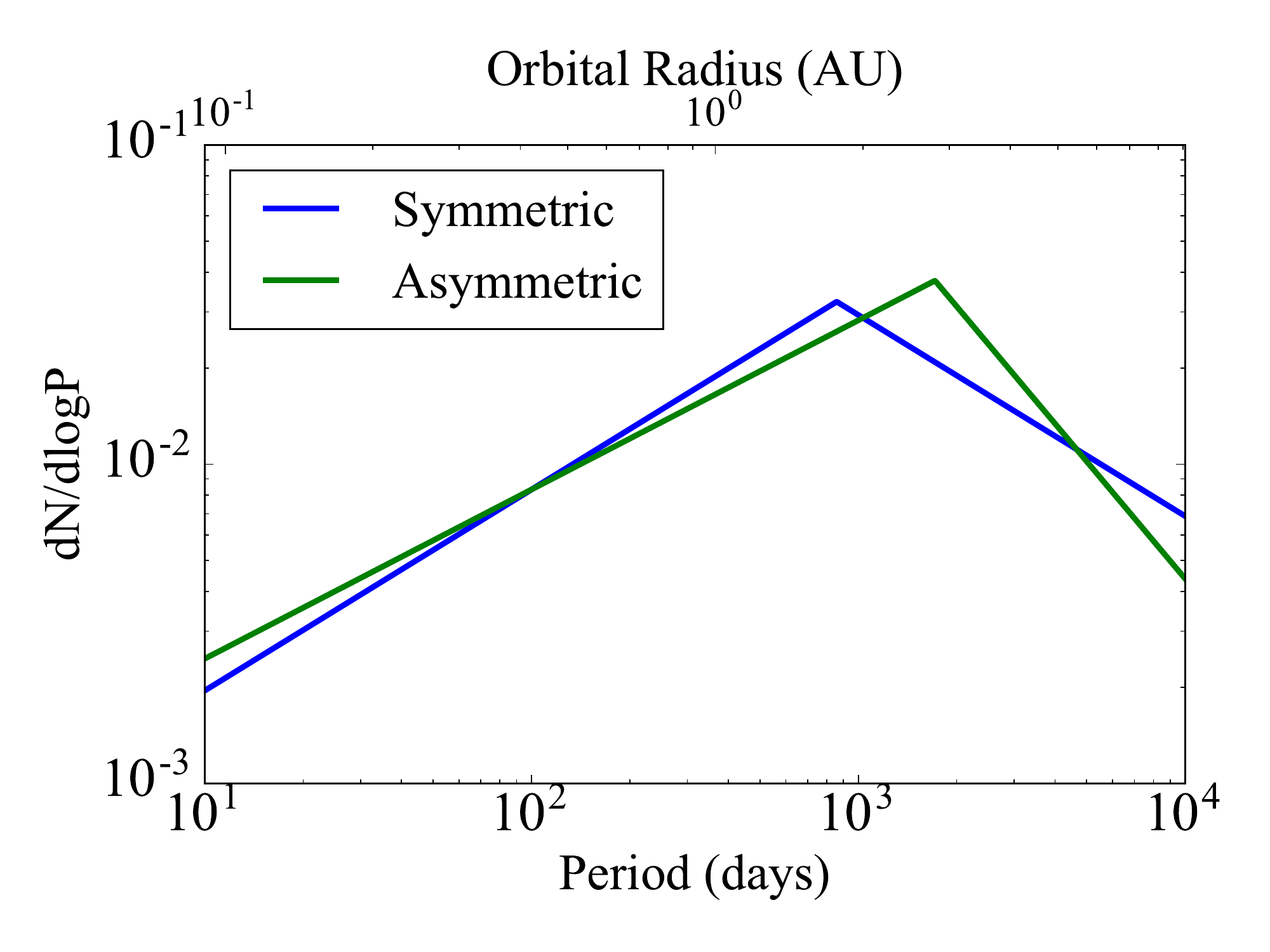}
\caption{Planet period distribution based on \citet{fernandes2019hints}, integrating over a mass range of 0.3--5\,$M_{J}$.}
\label{fig:dist}
\end{figure}

If we integrate over the mass range, Figure~\ref{fig:dist} shows the symmetric and asymmetric distributions of periods and orbital radii, similar to Figure 3 in \citet{fernandes2019hints}.
In this simulation, an optimistic orbital radius range of 0.6--60\,AU is considered.  If we integrate this function over our mass/orbital radius range, we obtain a total of 0.0808 and 0.0822 planets per star for the symmetric and asymmetric cases respectively.  This means, for the symmetric distribution, we expect 8.08\% of stars to have a planet in our mass/orbital radius range and 8.22\% for the asymmetric case.

\begin{figure}
\centering
\includegraphics[width=8cm]{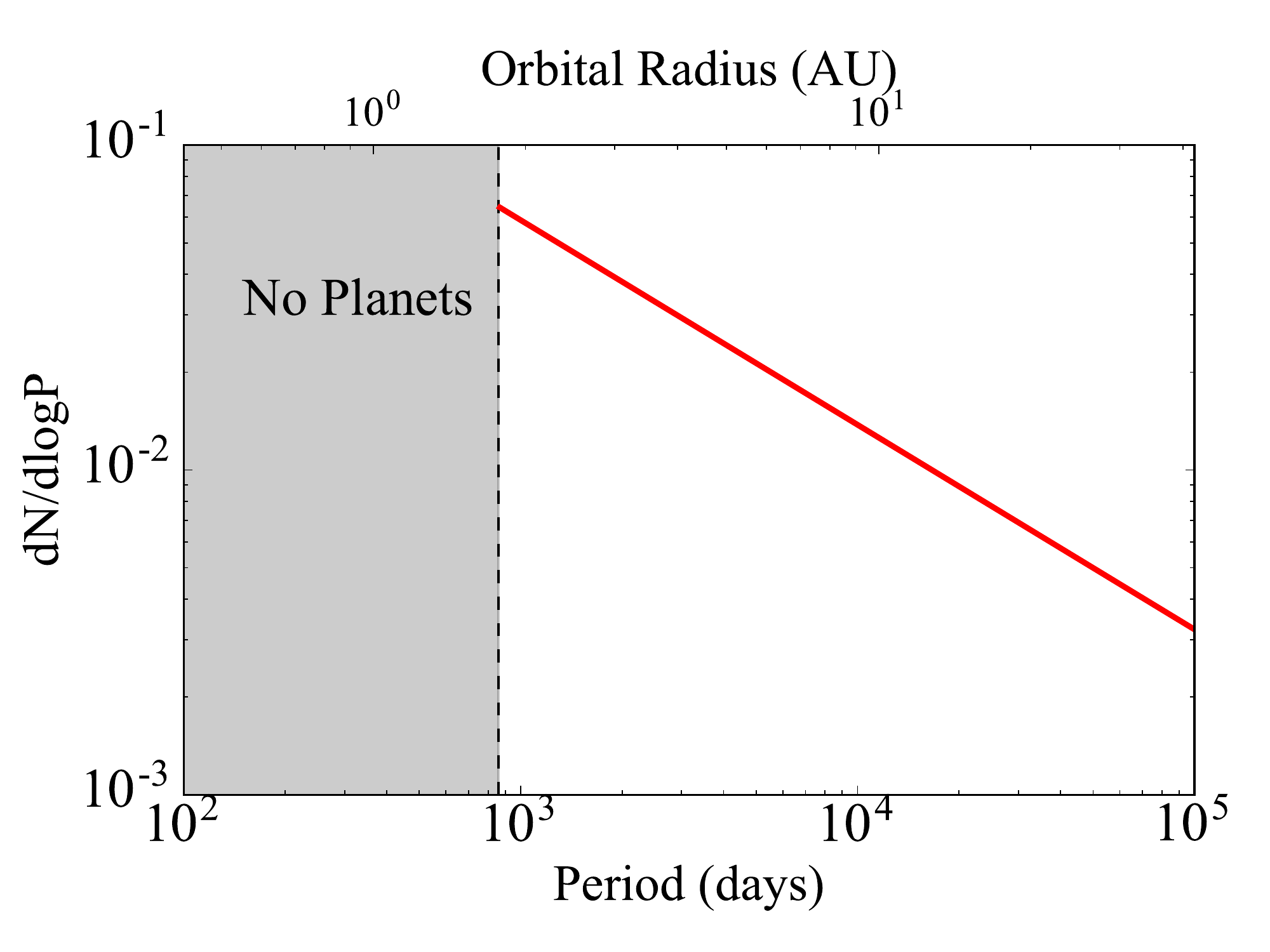}
\caption{Planet period distribution assuming no planets form within 1.77\,AU, integrating over a mass range of 0.3--5\,$M_{J}$.}
\label{fig:distMigration}
\end{figure}

However, these distribution do not take into account planet migration.  If we use these distributions, we are assuming the planets formed at their current positions which may not be the case.  For this reason, we also consider an extreme third distribution: a strong migration scenario in which no planets form at radii interior to the turnover point in the distribution from \citet{fernandes2019hints}, with all planets at small radii due to subsequent migration.  To create this distribution, we start with the symmetric distribution, set the occurrence rate to 0 for radii less than 1.77\,AU and double the normalization constant so the integral across all space will be the same.  The resultant distribution is shown in Figure~\ref{fig:distMigration}.
Using our planet distribution, we can randomly assign final masses and orbital radii to a sample of simulated planets.  In order to figure out how bright they will be, we simulate their formation by core accretion.
\section{A Model for Planet Luminosity with Core Accretion}
\label{sec:acc}
In the core accretion model of planet formation, a planet starts forming in a proto-planetary disc by accumulating small particles of dust and rock, first by Van der Waals interactions, then by gravitational interactions mediated by momentum exchange with the gas.
If the proto-planet accumulates sufficient mass ($\sim 10\,M_{\earth}$, \citet{pollack1996formation}) it can begin accreting gas at a rate sufficient to induce runaway growth.

For this simulation our main input parameters are:
\begin{itemize}
\item Final Planet Mass and Orbital Radius sampled from a distribution given in \S\ref{sec:dist}
\item The gas surface density of the disc as a function of orbital radius and time
\item The post-shock entropy of the planet
\end{itemize}
\noindent Note that this model ignores planet migration which may have a significant effect, especially when the disc is massive.  However, we assume here that the majority of the accretion luminosity for massive planets occurs in the later stages when the disc is close to dissipating. At this point, although there is high gas density near the accreting planet, the disc itself is not sufficiently massive to enable substantial migration of the most massive (and detectable) planet in each system.  We assume all migration happens after the end of our models.

We do not include solid accretion in our calculations. This is likely to have important observational signatures at early stages of giant planet formation, where protoplanetary discs are more likely to be optically thick. There have been suggestions, however \citep{van2017three} that an instrument such as METIS is capable of detecting the signatures of solid accretion in low mass cores.
\subsection{Disc Properties}
The initial surface density as a function of orbital radius is given by:
\begin{equation}
\Sigma_{0} = kr_{\rm orb}^{p}
\label{eq:densityGeneral}
\end{equation}
where we take $p\sim -3/2$ \citep[e.g][]{williams2011protoplanetary} and $k$ is a constant calculated from the total disc mass and outer radius.  In the simulation, the disc is given a total mass (20\,M$_{\rm{J}}$) and an outer radius set to a typical outer radius of 60\,AU from \citet{tripathi2017millimeter}.
The constant $k$, and hence the surface density at a given radius, is calculated by integrating Equation~\ref{eq:densityGeneral} over the disc surface area to obtain the total mass.  The constant $k$ is therefore given by:
\begin{equation}
k = \frac{(p+2)M_{D}}{2\pi R_{D}^{p+2}}
\end{equation}
where $M_{D}$ is the disc gas mass and $R_{D}$ is the disc radius.  If the power $p$ is -3/2, this gives an initial density of:
\begin{equation}
\Sigma_{0} = \frac{M_{D}}{4\pi \sqrt{R_{D}}}r_{\rm orb}^{-3/2}
\end{equation}

This gives a density of 1743 gcm$^{-2}$ at 1AU which is close to the value for the minimum mass solar nebula (MMSN) given in \citet{hayashi1981structure}.  Similar to \citet{lissauer2009models}, the disc is expected to dissipate over time and a linear decline in density is assumed:
\begin{equation}
\Sigma(t)=-\frac{\Sigma_{0}t}{\tau_{D}}
\end{equation}
where $\Sigma_{0}$ is the initial density and $\tau_{D}$ is the maximum time for gas to accrete (3\,Myr in this simulation) and $-\tau_{D}<t<0$.  This surface density evolution is the most optimistic case.  A parabolic evolution was attempted, in which the density drops quickly at the beginning before slowing down.  This was found to reduce the planet brightness by $\sim$0.5 magnitudes during accretion and have almost no effect on the post formation magnitude.  For this reason, we do not consider any other surface density evolution functions and all results in this paper use the linear decline.

The protoplanetary disc temperature follows a power law. We use the optically thick, passively heated, flared disc power law from \citep{chiang2010forming}:
\begin{equation}
T(r_{\rm{orb}}) = 120\mathrm{K}\left(\frac{r_{\rm{orb}}}{1\mathrm{AU}}\right)^{-3/7}
\end{equation}
This is the temperature at the midplane of the disc.  In this simulation we assume all planets form at the midplane so we do not consider the vertical temperature profile.  The midplane temperature is then used to calculate the isothermal sound speed given by
\begin{equation}
c_{s} = \sqrt{\frac{k_{B}T}{\mu m_{H}}}
\end{equation}
where $k_{B}$ is the Boltzmann constant, $\mu$ is the mean molecular weight and $m_{H}$ is the mass of a hydrogen atom.  The value of $\mu$ is taken from \citet{kimura2016birth} where $\mu=2.35$.  Additionally, we can calculate the Toomre $Q$ parameter given by
\begin{equation}
Q = \frac{c_{s}\kappa}{\pi G\Sigma}
\end{equation}
where $\kappa$ is the epicyclic frequency.  Assuming a Keplerian disc, this is equal to the angular speed.  The stability criterion for the disc is $Q>1$.  For our disc, $Q=35.5$ at 1\,AU and 14.8 at 60\,AU (the outer edge of the disc.)  Since this disc is stable against gravitational collapse, we only consider planet formation by core accretion.

Table~\ref{tab:properties} summarizes the properties of the disc and the values we use in this simulation.\\
\begin{table}
\begin{tabular}{l|l|l}
Parameter & Meaning & Value\\\hline
$M_{D}$ & Total Gas Mass & 20 M$_{\rm{J}}$\\
$R_{D}$ & Disc Radius & 60\,AU\\
$\tau_{D}$ & Gas Accretion Time & 3\,Myr\\
$M_{*}$ & Stellar Mass & 1$M_{\odot}$\\
$T$ & Disc Gas Temperature & $120\mathrm{K}\left(\frac{r_{\rm{orb}}}{1\mathrm{AU}}\right)^{-3/7}$\\\\
$\Sigma_{0}$ & Initial Gas Surface Density & $\frac{M_{D}}{4\pi \sqrt{R_{D}}}r_{\rm orb}^{-3/2}$\\\\
$\Omega$ & Keplerian Frequency & $\frac{GM_{*}}{r_{\rm orb}^{3}}$\\\\
$c_{s}$ & Isothermal Sound Speed & $\sqrt{\frac{k_{B}T}{\mu m_{H}}}$\\\\
$H$ & Disc Thickness & $c_{s}/\Omega$\\
\end{tabular}
\caption{Summary of Disc Properties}
\label{tab:properties}
\end{table}
\subsection{Growth of Thermally Regulated Envelope}
In the early stages of gas accretion, the planet consists of a dense solid core surrounded by an envelope of gas embedded in the proto-planetary disc.  This envelope starts off optically thin, however, as more gas is added to the envelope, it becomes extremely optically thick which prevents most emitted radiation from escaping to space.  This causes the envelope's density and temperature to greatly exceed that of the surrounding disc \citep{d2010giant}.

The resultant pressure gradient opposes the gravitational attraction preventing the accretion of large amounts of gas.  During this time, the radius of the planet is poorly defined but we use the accretion radius given by equation 3 in \citet{lissauer2009models}.  This radius is quite large ($\sim 50R_{J}$) and the planet would be undetectable during this time at near- and mid-infrared wavelengths.

During this time, we represent the planet's accretion rate with a version of a power law derived by \citet{ikoma2000formation}:
\begin{equation}
\frac{dM}{dt}=\frac{1}{\bar{\tau}_{e}}\frac{M^{\xi +1}}{M_{\earth}^{\xi}}
\label{eq:mdotFirst}
\end{equation}
The constants $\xi$ and $\bar{\tau}_{e}$ depend on a number of factors such as envelope opacity and temperature.  The models in \citet{d2010giant} use $\xi$=3 and $\bar{\tau}_{e}=10^{10}$ years but $\bar{\tau}_{e}$ can vary significantly depending on gas  parameters.

In this simulation, $\xi$ is fixed at 3 and $\bar{\tau}_{e}$=10$^{a}$ years where $a$ varies between 5 and 10.  The resultant value of $\bar{\tau}_{e}$ determines the maximum mass a planet can attain before the disc dissipates.  Once a planet is assigned a final mass and orbital radius, $\bar{\tau}_{e}$ is adjusted until this final mass is equal to the maximum mass.  This ensures a physical situation in which the planet finishes accreting once the disc dissipates.  A value of $\bar{\tau}_{e}$=10$^{5}$ years may seem extremely small but this is only the minimum allowed in our simulation and will only occur for the most massive planets at very wide separations which will be extremely rare. Alternative initial conditions with a more massive but still Toomre stable disc would significantly increase this required time.
\subsection{Runaway Accretion}
Once the gaseous envelope reaches a critical mass, it becomes gravitationally unstable and contracts onto the planet.  The exact value of this critical mass is a matter of debate but is assumed to be when the envelope mass is comparable to the core mass \citep{mordasini2007giant}.  In this work, we assume the critical envelope mass is 4/3 times the core mass as shown in \citet{d2010giant}.  
This begins the phase known as runaway accretion.  During this phase, gas rapidly accretes onto the planet causing the envelope to again increase in mass and contract.  This increase in the planet's mass causes the accretion rate to increase leading to more gas capture.  During this rapid growth, the planet will start to clear a gap in the disc, becoming visible at most system inclinations, and reach its maximum luminosity.

\begin{figure}
\centering
\includegraphics[width=8cm]{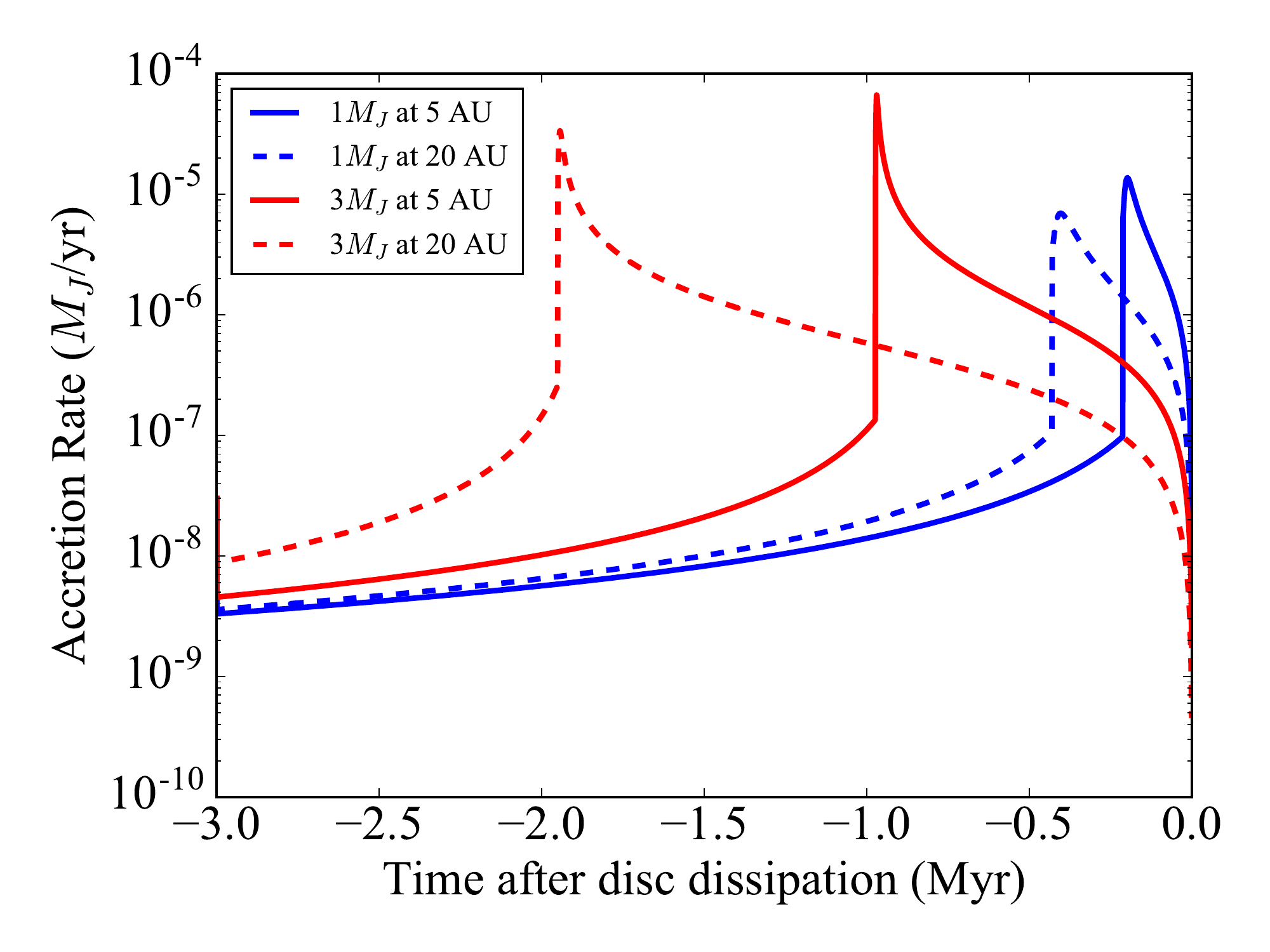}
\caption{Accretion Rate as a function of time for 1 and 3\,M$_{\rm{J}}$ planets at 5 and 20\,AU.  The initial phase of accretion is shorter for the 3\,M$_{\rm{J}}$ planets.  This allows them to spend more time on the runaway accretion (after the spike) and attain their target mass of 3\,M$_{\rm{J}}$.  All planets finish accreting when the disc dissipates.}
\label{fig:mdot1}
\end{figure}

In this stage of formation, the planet's accretion rate strongly depends on the proto-planetary disc properties and the planet's radius of gas capture.  From hydrodynamical simulations, the accretion rate can be approximated by
\begin{equation}
\frac{dM}{dt} \sim \frac{\Sigma}{H}\Omega R_{\mathrm{gc}}^{3}
\label{eq:accRate1}
\end{equation}
where $\Sigma$ is the disc gas surface density, $\Omega$ is the Keplerian frequency, $H$ is the local disc thickness defined in table~\ref{tab:properties} and $R_{\mathrm{gc}}$ is the radius of gas capture \citep{d2010giant}.  The value of $R_{\mathrm{gc}}$ is defined as the smaller value between the Hill radius ($R_{H}$) and the Bondi radius ($R_{B}$) given by
\begin{equation}
R_{H} = r_{orb}\left(\frac{M_{p}}{3M_{\star}}\right)^{1/3}
\end{equation}
and
\begin{equation}
R_{B} = \frac{GM}{c_{s}^{2}}
\end{equation}
\noindent The planet continues to grow until there is no more gas to accrete which happens either when the planet has cleared a wide enough gap or the disc has dissipated.  In this simulation, the accretion rate is calculated from a viscosity-limited log-parabolic relation given in \citet{lissauer2009models}:
\begin{equation}
\mathrm{log}_{10}\left(\frac{\dot{M}}{\Sigma r^{2}/P}\right)=c_{0}+c_{1}\mathrm{log}_{10}(\mathrm{M/M}_{\mathrm{J}})+c_{2}\mathrm{log}^{2}_{10}(\mathrm{M/M}_{\mathrm{J}})
\end{equation}
where $r$ is the planet's orbital radius and $P$ is its period.  Note this parabolic fit is independent of $\Sigma$, $r$ and $P$ and depends solely on the planet mass.  The coefficients depend on the disc viscosity.  For this simulation, the high viscosity case (viscosity parameter $\alpha = 4\times 10^{-3}$)
is chosen giving the following values:  $c_{0} = -2.87$, $c_{1} = -1.63$ and $c_{2} = -1.28$.  We use this parabolic fit in our model rather than Equation~\ref{eq:accRate1}, which is only included for reference, in order to be as simplistic as possible.  In this model, the gas accretion rate increases as the planet becomes more massive but then decreases as a gap forms in the disc.  This dimensionless accretion rate is made physical after multiplying by $\Sigma r^{2}/P$.  This causes the accretion rate to decline even faster as $\Sigma$ is decreasing with time.  Figure~\ref{fig:mdot1} shows the accretion rates for two different planet masses at two different separations.  The disc has a mass of 20\,M$_{\rm{J}}$ and radius 60\,AU.  For simplicity, we assume gas accretion starts exactly 3 million years before the disc dissipates for all planets.

As shown in Figure~\ref{fig:mdot1}, out of the four examples, the runaway accretion phase is the shortest for a 1\,M$_{\rm{J}}$ planet at 5\,AU, only lasting the final 200,000 years before the disc dissipates.  If the planet is placed further away, the runaway accretion is slower due to lower gas density in the disc.  A higher mass planet (3\,M$_{\rm{J}}$) requires more time in the runaway accretion phase.  Since our planet masses are based on observed distributions and not disc properties, we assume the disc has enough time to form these massive planets.  In order to achieve this, higher mass planets must begin the runaway accretion phase earlier as can be seen in Figure~\ref{fig:mdot1}.  All planets finish accreting when the disc dissipates which we define as $t=0$.\\

\noindent To determine a planet's luminosity during the accretion phase, we need its mass, accretion rate and radius.  During the runaway accretion phase, the radius is calculated from the planet's entropy.
\subsection{Planet Entropy}
In order to understand the entropy of an accreting planet, we used Modules for Experiments in Stellar Astrophysics (MESA) \citep{paxton2010modules,paxton2013modules,paxton2015modules}.  The \texttt{make\_planet} test suite in MESA was used to model planets of different sizes and masses and extract the entropy of the internal convective zone.

\begin{figure}
\centering
\includegraphics[width=10cm]{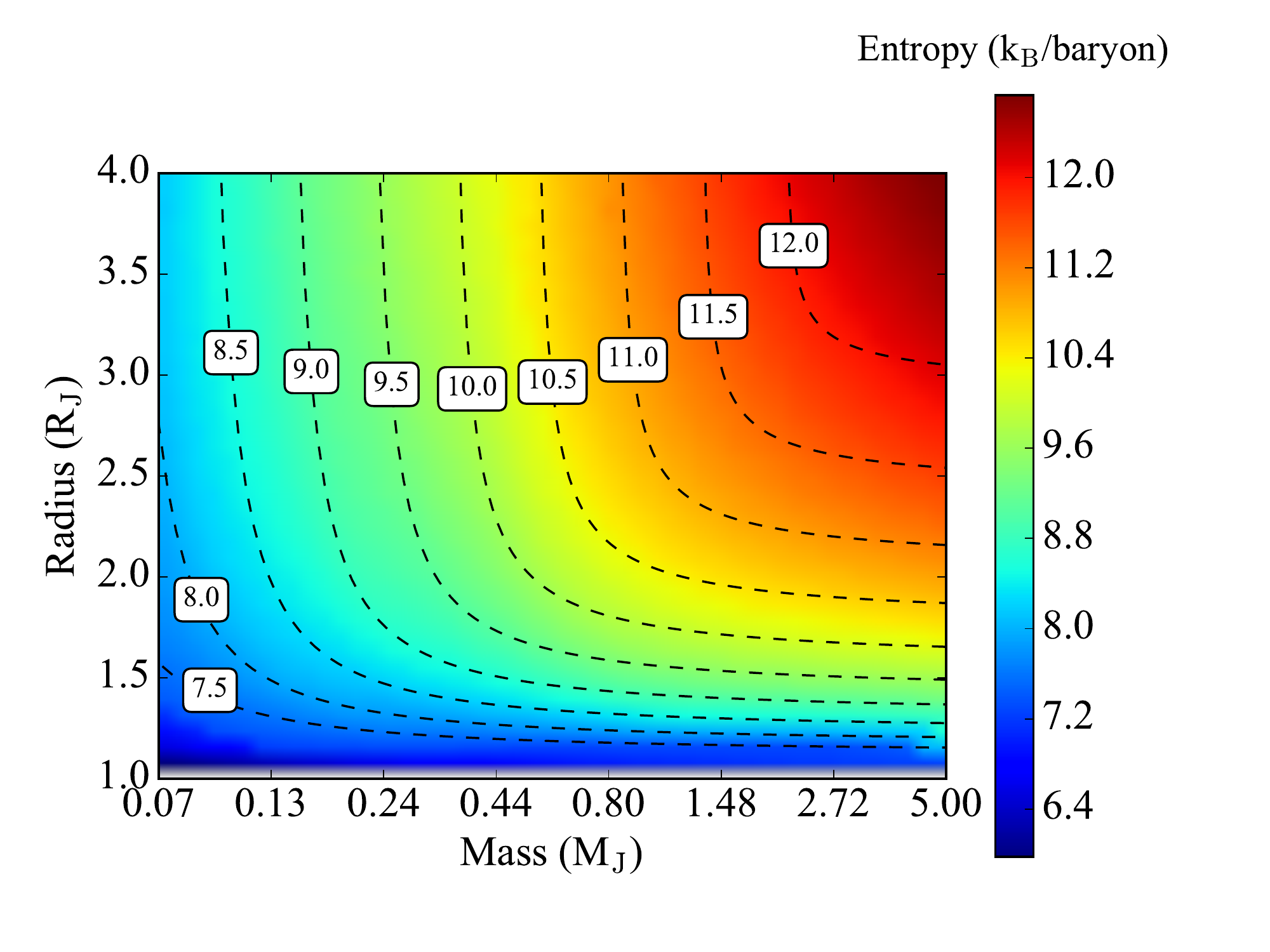}
\caption{Entropy as a function of mass and radius for a core of mass 10\,M$_{\earth}$ taken from MESA.  The lines of constant entropy approximately follow an inverse power law.  If we assume a constant entropy in the center, the planet should shrink as its mass increases.}
\label{fig:sPlot}
\end{figure}

Planets were modeled over a mass range of 0.07--5\,M$_\mathrm{J}$ and radius range of 1--4\,R$_{\mathrm{J}}$.  For all planets modeled, we assume a core of 10\,M${\earth}$ with a density of 10\,g/cm$^{3}$ and hence radius of $\sim$ 1.76\,R${\earth}$.  All planets have Solar metallicities, elemental abundances taken from \citet{grevesse1998standard} and opacities from \citet{freedman2008line}.  MESA uses the equation of state from \citet{saumon1995equation} and the atmospheric model we used was \texttt{simple\_photosphere}.  We tried modifying the atmospheric model to \texttt{tau\_10\_tables} but this produced the same result for planet entropy to within 0.5\%.  The entropy as a function of planet mass and radius is shown in Figure~\ref{fig:sPlot}.
The `\texttt{inlist}' files containing our parameters can be found at \url{https://github.com/awallace142857/planet_simulator}.  These files are for a planet with mass 1\,M$_{\rm{J}}$ and radius 1.5\,R$_{\rm{J}}$.  The files for other masses and radii contain the same parameters with only the total mass and radius changed.

In order to determine how the planet's internal entropy changes over time, we note there is a characteristic post-shock temperature $T_{c}$ given by:
\begin{equation}
    4\pi R^{2}\sigma_{SB}T_{c}^{4}\sim \frac{GM\dot{M}}{R}
\end{equation}
For the mass range considered in this simulation, the planet accretes the majority of its mass after envelope collapse at a small radius ($\sim$ 1.5$R_{J}$.)  The minimum accretion rate after envelope collapse is approximately $1\times 10^{-7}$ $M_{J}$/yr.  Using the minimum planet mass in our simulation (0.3$M_{J}$,) we obtain $T_{c}\sim$700\,K.  If the post-shock temperature is significantly below this value, then the post shock gas has to radiatively cool through hot gas that has a blackbody flux that exceeds its own blackbody emission. This is certainly possible if e.g. dominated by line opacity that is a strong function of temperature, but we take the position that this is an unlikely scenario.  Recent work by \citet{marleau2017planetary} found that for accretion rates and planet masses similar to those considered here, the post shock temperature is above 1000K.  This places our planets well within the stalling regime and on the edge of the cooling regime in \citet{berardo2017evolution}.  From this we conclude that, for our chosen entropies, the internal entropy changes little during the accretion phase \citep{marleau2019planetary}.

If we assume a constant entropy after envelope collapse, the lines of constant entropy in Figure~\ref{fig:sPlot} can be used to calculate how the radius shrinks with increased mass.  Using our minimum value for $T_{0}$ and Figure 6 from \citet{berardo2017evolution}, we determine a lower entropy limit of $\sim$9.5$k_{\rm{B}}$/baryon.  For lower mass planets accreting more slowly, lower entropies are still possible but these planets are not considered due to low observation sensitivities.  Should a significantly more sensitive (i.e. space-based) instrument be considered in a future study, these lower entropies should also be considered.

A more realistic scenario might be one with lower initial entropy, but hot accretion at very early stages, resulting in a stalling accretion \citep{berardo2017evolution} at intermediate entropy values similar to our assumed 9.5-10.5$k_{\rm{B}}$/baryon while the bulk of the planet mass is accreted.  In this paper, we assume two entropies at the extremes of the stalling regime: 9.5$k_{\rm{B}}$/baryon (cold extreme) and 10.5$k_{\rm{B}}$/baryon (hot extreme.)

\begin{figure}
\centering
\includegraphics[width=8cm]{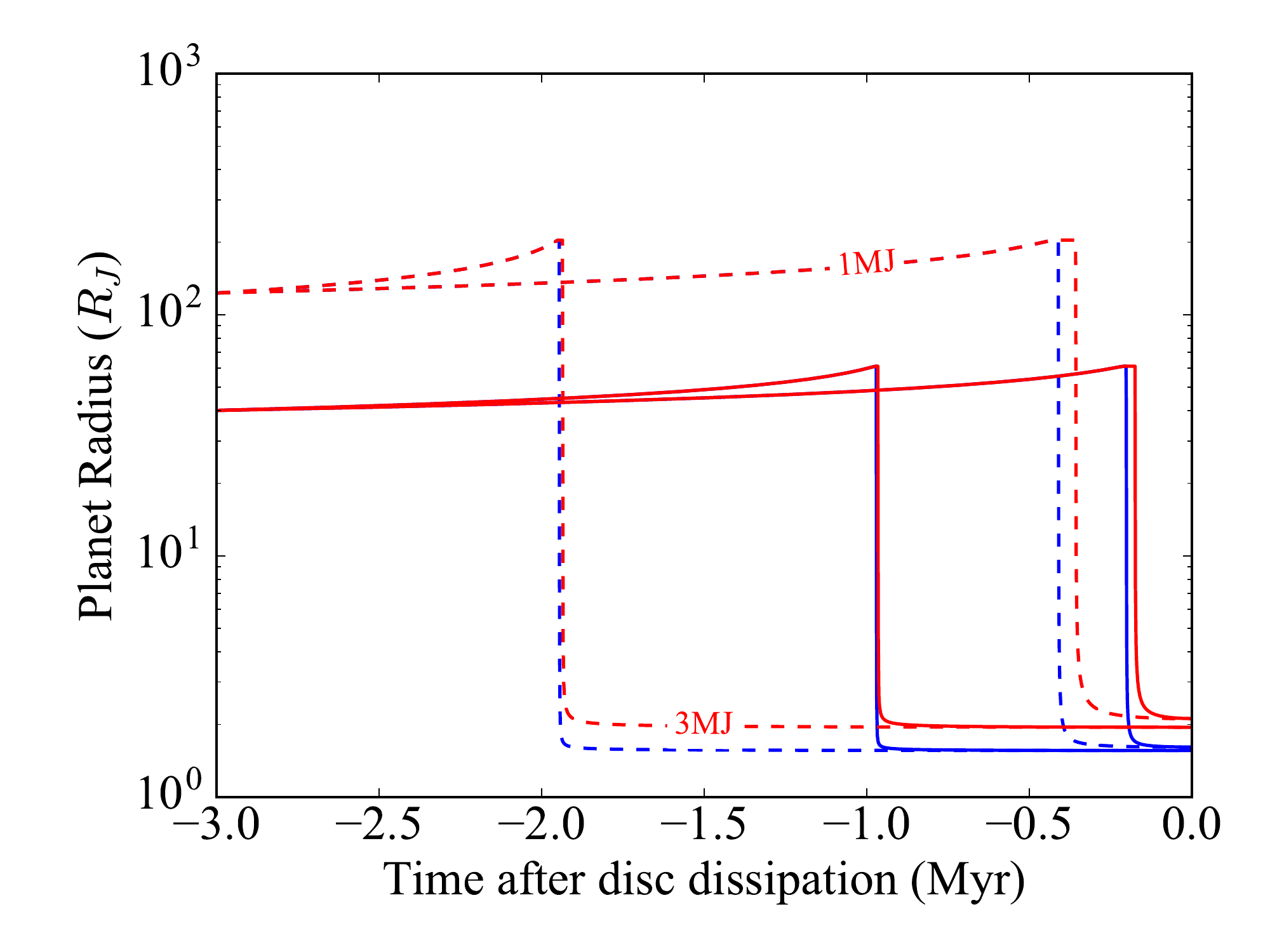}
\caption{Planet radius as a function of time for 1 and 3\,M$_{\rm{J}}$ planets at 5 and 20\,AU.  \textbf{Solid lines:} Planets at 5\,AU.  \textbf{Dashed lines:} Planets at 20\,AU.  \textbf{Blue:} Entropy = 9.5$k_{\rm{B}}$/baryon.  \textbf{Red:} Entropy = 10.5$k_{\rm{B}}$/baryon.  In the initial stages of accretion, the radius is determined solely by the distance from the star.  The 20\,AU planets are larger because they have a larger Hill radius.  During runaway accretion, the radius decreases according to the constant entropy curves in Figure~\ref{fig:sPlot}.}
\label{fig:radPlot}
\end{figure}

Figure~\ref{fig:radPlot} shows the radius evolution for the same planets shown in Figure~\ref{fig:mdot1} assuming internal entropies of 9.5$k_{\rm{B}}$/baryon (blue) and 10.5$k_{\rm{B}}$/baryon (red) and fitting an inverse power law (dashed curves in Figure~\ref{fig:sPlot}.)  These entropies are only used during and after runaway accretion when the planet is assumed to be fully convective.  In the early phase of accretion, the planet radius is assumed to be the accretion radius.

As shown in Figure~\ref{fig:radPlot}, the planet radius increases to approximately 60 R$_{\rm{J}}$ as the envelope grows, then rapidly shrinks down below 3 R$_{\rm{J}}$ when the envelope collapses.  As the planet continues accreting, its internal entropy remains constant and the planet contracts according to the curves of constant entropy shown in Figure~\ref{fig:sPlot}.  The higher entropy planets result in a larger final radius.  Using this radius value we can now calculate the planet's luminosity.
\subsection{Planet Luminosity}
During the accretion phase (both before and during runaway accretion,) the planet's luminosity is calculated according to models from \citet{zhu2015accreting} where the bolometric luminosity of the accretion disc is given by
\begin{equation}
L = \frac{GM\dot{M}}{2R_{d}},
\label{eq:lumbol}
\end{equation}
where the inner radius of the disc $R_d$ is set to the radius of the planet $R_p$ in our calculations, corresponding to boundary layer accretion.  This equation only refers to half of the total energy transfer during accretion.  We conservatively assume that the other half of the luminosity is lost in the accretion shock at non-observable wavelengths.

\begin{figure}
\centering
\includegraphics[width=8cm]{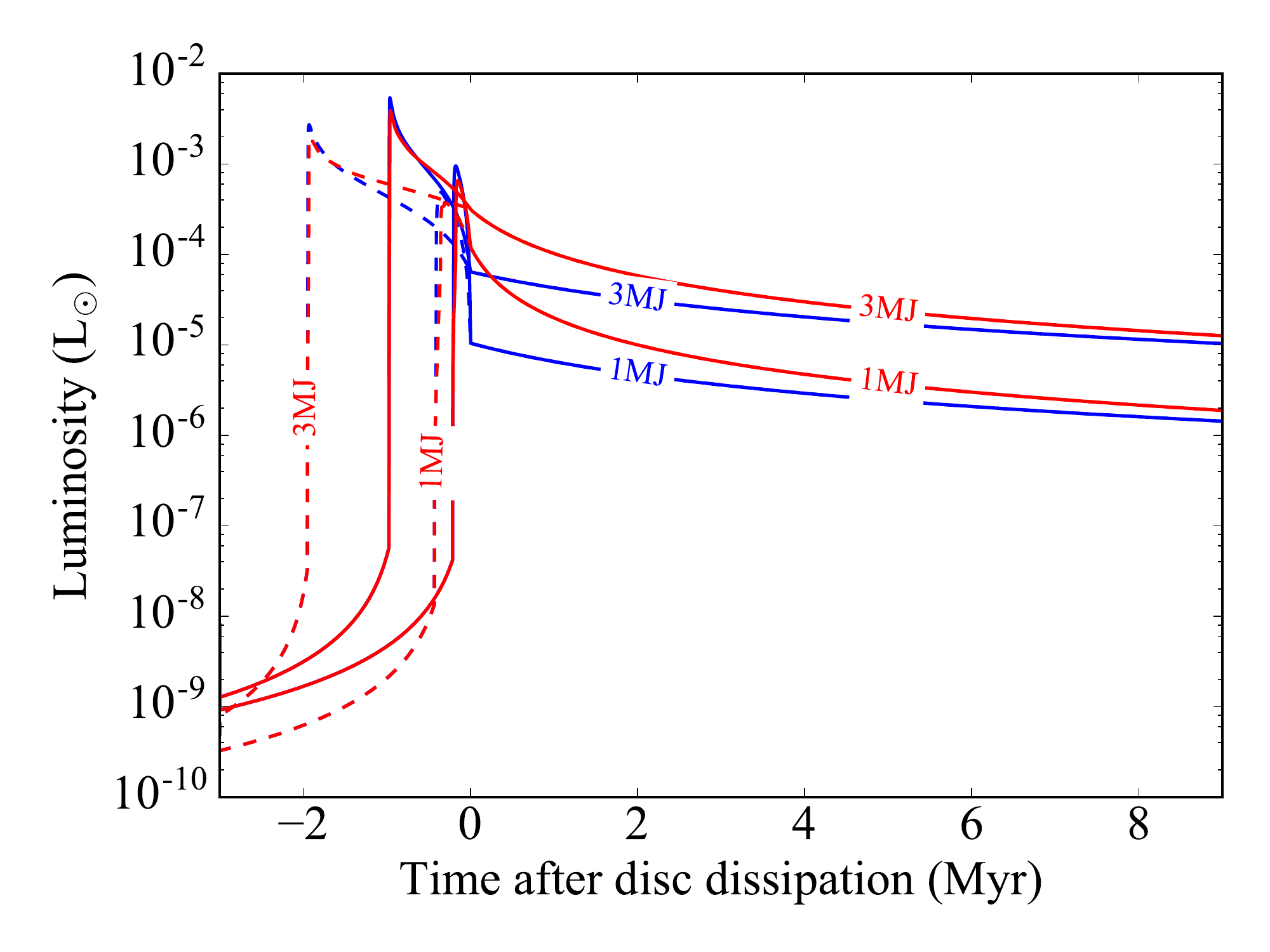}
\caption{Planet luminosity as a function of time for 1 and 3\,M$_{\rm{J}}$ planets at 5 and 20\,AU.  \textbf{Solid lines:} Planets at 5\,AU.  \textbf{Dashed lines:} Planets at 20\,AU.  \textbf{Blue:} Entropy = 9.5$k_{\rm{B}}$/baryon.  \textbf{Red:} Entropy = 10.5$k_{\rm{B}}$/baryon.  The 3\,M$_{\rm{J}}$ planet at 5\,AU is the brightest due to it having the highest mass and accretion rate.  All planets finish accreting at $t=0$ and the post-accretion luminosity depends solely on mass and radius.  In the final stages of accretion and after formation, the higher entropy planets are brighter due to their larger radii.}
\label{fig:lumbolAll}
\end{figure}

Using the planet's mass and radius, the post-accretion luminosity is calculated by MESA using the same abundances, opacities and equation of state as before.  During the final stages of accretion, the luminosity from \citet{zhu2015accreting} is combined with this post-accretion luminosity at an age of 0 to provide a more smooth transition from accreting to non-accreting.
Using the example planets shown in Figures~\ref{fig:mdot1} and~\ref{fig:radPlot} with the same internal entropies, Figure~\ref{fig:lumbolAll} shows how each planet's luminosity changes during and after the accretion phase.

As shown in Figure~\ref{fig:lumbolAll}, during the runaway accretion phase, the planets are at their brightest  with luminosities of $\sim 10^{-3}L_{\odot}$.  The 3M$_{\rm J}$ planet at 5\,AU is the brightest due to its fast accretion rate and high mass.  The large increase in luminosity is mainly due to the increased accretion rate shown in Figure~\ref{fig:mdot1} but also due to the reduced radius calculated from the trend in Figure~\ref{fig:sPlot}.  The luminosity then declines as the accretion slows down.  Due to their larger radii, the higher entropy planets are initially fainter in the early stages of accretion, as per Equation~\ref{eq:lumbol} but then become brighter in the final stages as the luminosity from the planet becomes dominant over accretion.

Once the planets have formed, they gradually cool and the luminosity decreases accordingly.  As shown in Figure~\ref{fig:lumbolAll}, the post-accretion luminosity is independent of orbital radius.  This is because we assume constant internal entropy for each planet which means two planets of equal mass will have equal radius regardless of their orbital radius.

\begin{figure}
\subfigure[L' magnitudes]{\label{fig:LmagEx}\includegraphics[width=8cm]{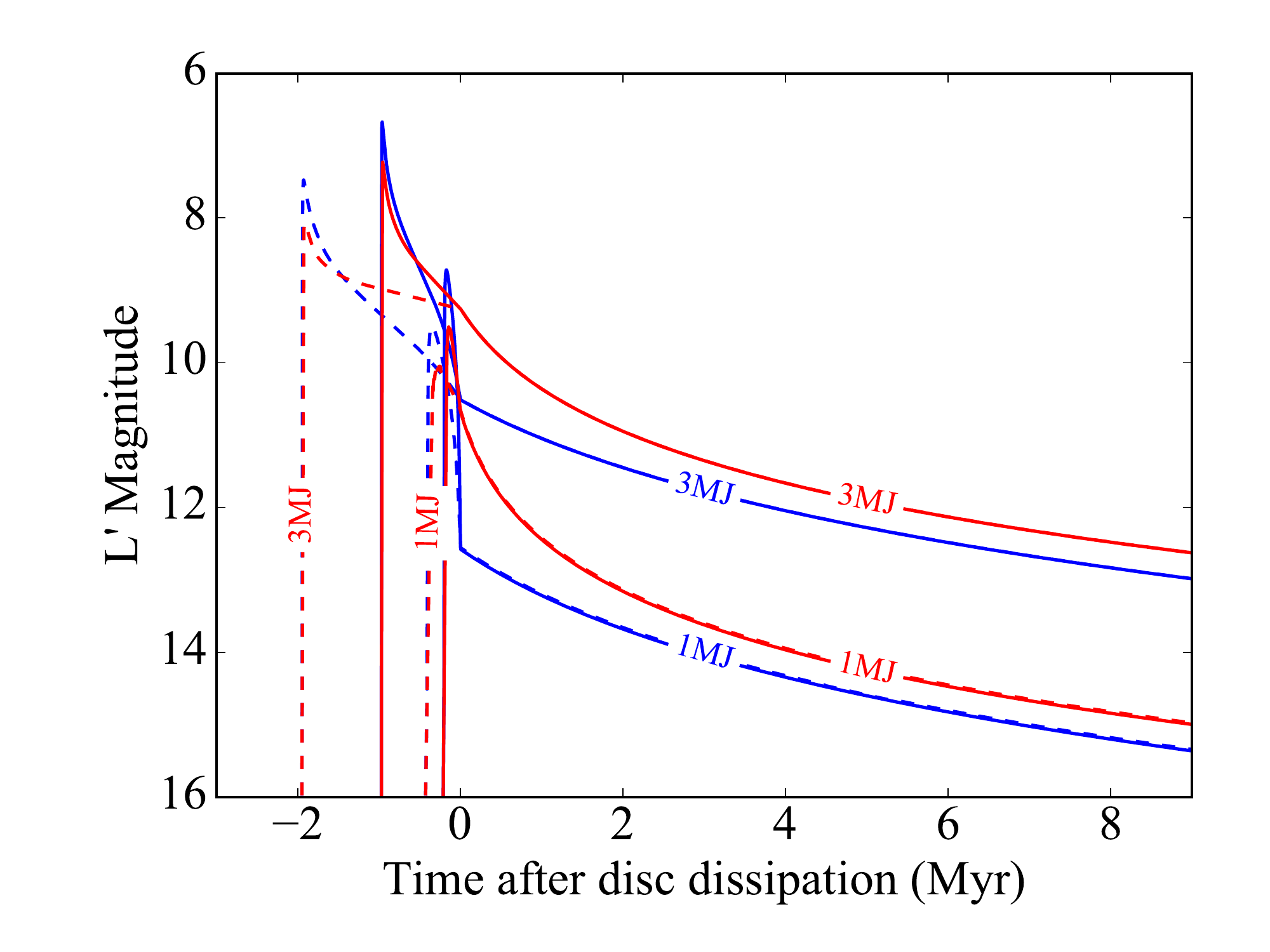}}
\subfigure[K magnitudes]{\label{fig:KmagEx}\includegraphics[width=8cm]{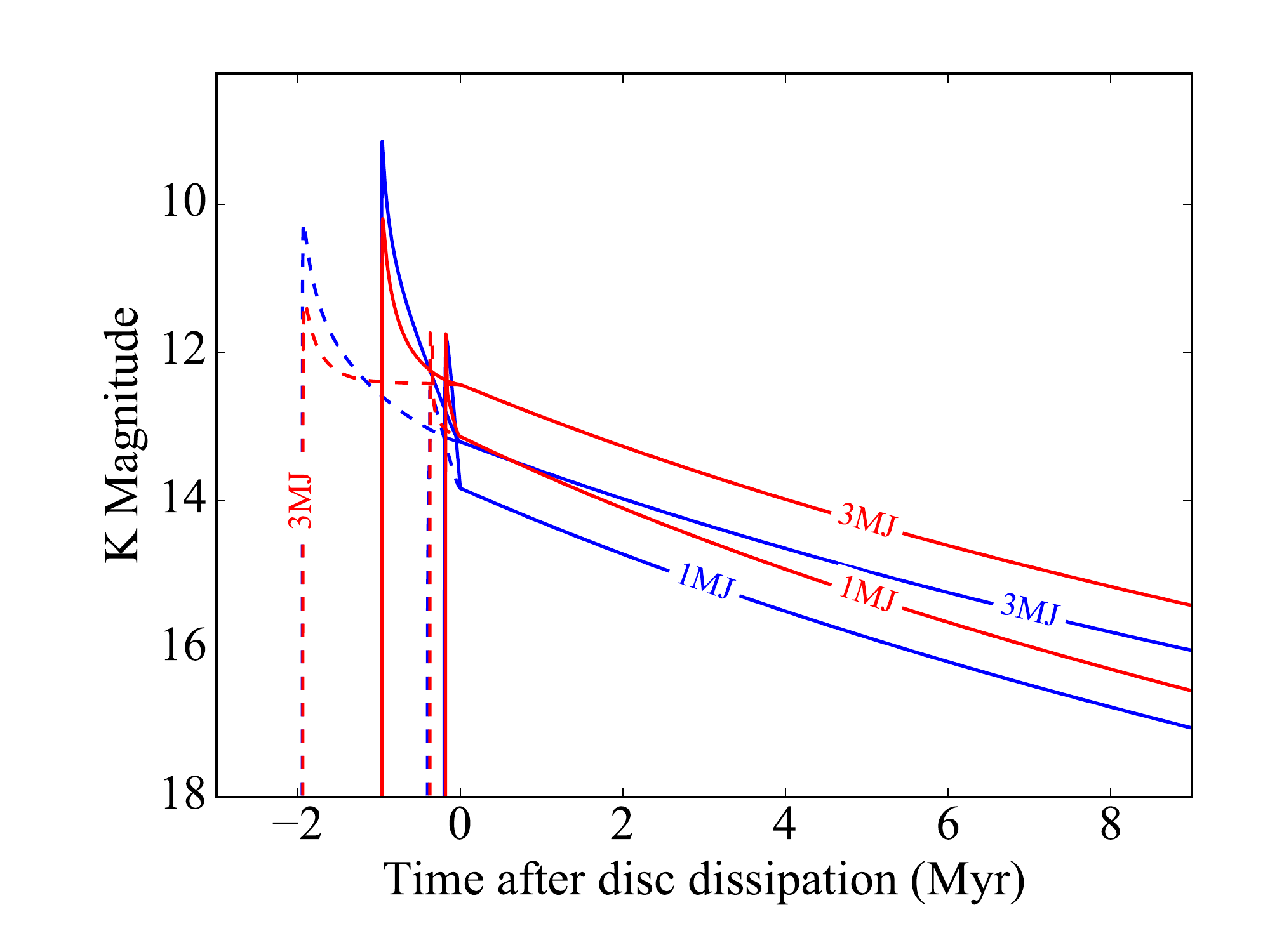}}
\caption{Magnitude as a function of time for different filters.  \textbf{Solid lines:} Planets at 5\,AU.  \textbf{Dashed lines:} Planets at 20\,AU.  \textbf{Blue:} Entropy = 9.5$k_{\rm{B}}$/baryon.  \textbf{Red:} Entropy = 10.5$k_{\rm{B}}$/baryon.  During the first stages of accretion, the planets are too faint to be shown on these plots.  They are all at their brightest during runaway accretion and grow fainter as accretion slows down.  Once they clear a gap in the disc, this would be the optimal time to find them.}
\label{fig:magEx}
\end{figure}

During formation, a planet's magnitude  is calculated for different filters using models from \citet{zhu2015accreting} which depends on the planet's mass, radius and accretion rate.  In the final stages of planet formation, this is combined with a hot-start magnitude from \citet{spiegel2012spectral} at an age of 0.  This magnitude depends only on the planet's mass and radius.  After the planet forms, the magnitude comes purely from \citet{spiegel2012spectral}. Figure~\ref{fig:magEx} shows the magnitude evolution of the example planets from Figure~\ref{fig:lumbolAll} in the L' and K filters which have central wavelengths of 3.75$\mu$m and 2.2$\mu$m respectively \citep{tokunaga2005mauna}.

From the magnitude plots in Figure~\ref{fig:magEx}, for most of the planet's formation, it is undetectable as it is growing very slowly and is still embedded in a protoplanetary disc.  During the runaway accretion phase, the massive planets approach
$M_{L'}\sim 8$, so should be easily detectable by even moderate contrast instruments.
However, this phase is very short and the $M_{L'}$ increases to 14 once the planet has formed.
The code used to simulate a large sample of forming planets can be found at \url{https://github.com/awallace142857/planet_simulator}.  This code also randomly assigns a planet mass and orbital radius from the distribution shown in Figure~\ref{fig:dist}.
\section{Likelihood of Planet Detection with Current and Future Instruments}
The simulation described in \S\ref{sec:acc} was run multiple times using different planet distributions from \S\ref{sec:dist} to obtain samples of planets and determine how many of these can be detected.  This is calculated for instruments operating in the L' and K bands assuming planet internal entropies of 9.5\,$k_{B}$/baryon and 10.5\,$k_{B}$/baryon.
\subsection{L'-band}
\begin{figure}
\centering
\includegraphics[width=8cm]{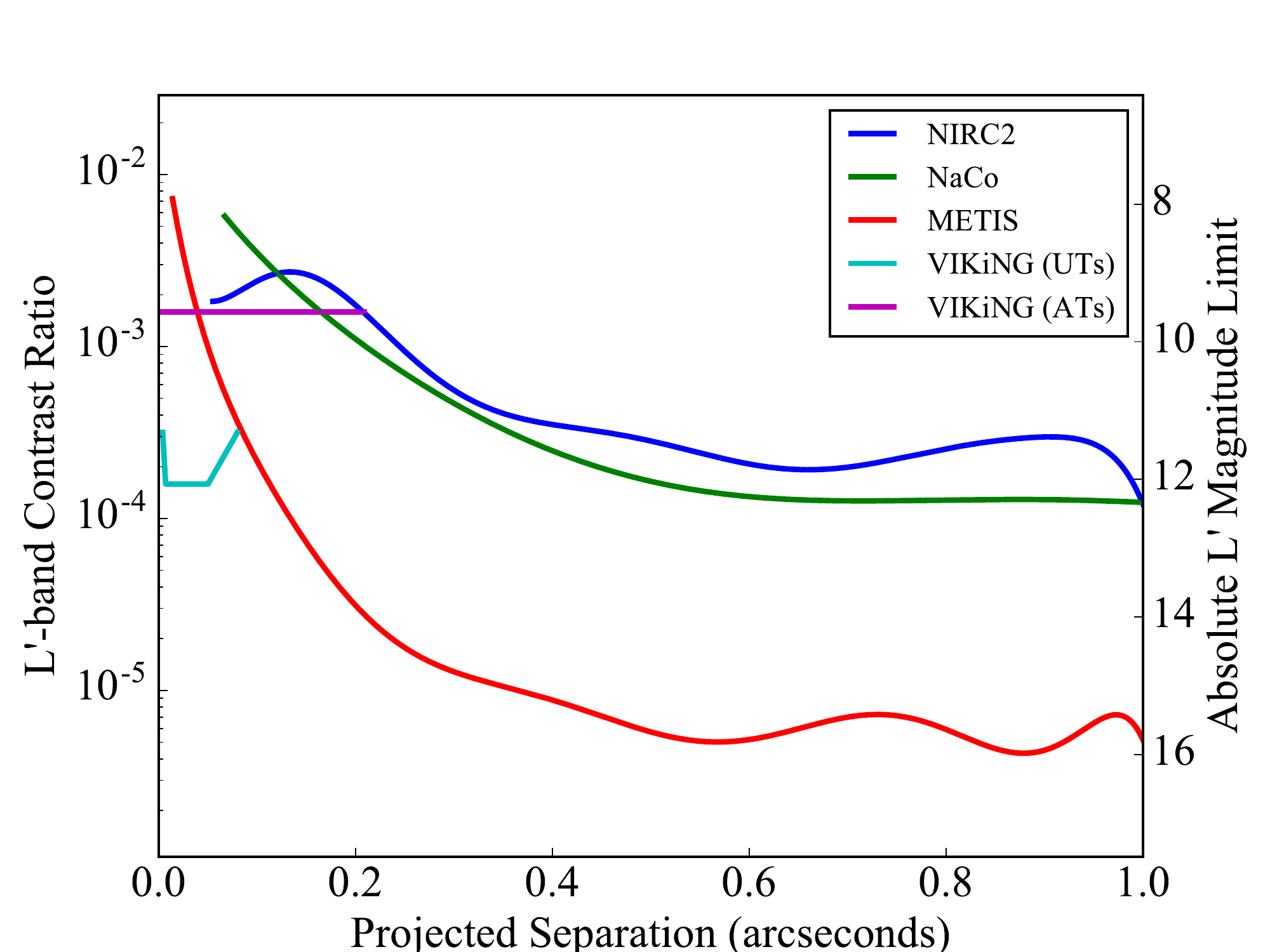}
\caption{L' contrast and magnitude limits for example instruments assuming a target of magnitude 8.3 at 140 pc.  For NIRC2, NaCo and METIS, these limits cut off at angles of 0.5$\lambda$/D which we assume to be the inner working angle without a coronagraph.}
\label{fig:limits}
\end{figure}

Figure~\ref{fig:limits} shows the limit of contrast ratio and L' magnitude for a  number of current and future instruments and techniques.  We assume a star in the Taurus molecular cloud which puts it at a distance of 140 pc \citep{kenyon1994new}.  The star we use in these examples is GM Aur which has an apparent L' magnitude of 8.3 (approximately average for Taurus.)  The limits for NIRC2 are based on our observations of Taurus (Wallace et al. 2019 in prep.), the NaCo limits are based on \citet{quanz2012searching} and the METIS limits are based on \citet{carlomagno2016end}.  We also examine the capabilities of VLTI Infrared Kernel NullinG (VIKiNG), a proposed instrument concept within the Hi-5 framework of the VLTI \citep{martinache2018kernel}.  For VIKiNG, the auxiliary telescopes (ATs) can be used to a field of view of 210 mas while the unit telescopes (UTs) can only be used out to 80 mas but have higher sensitivity.  An off-axis mode of Hi-5 is not considered but would have a similar background limited sensitivity at large radii.

As shown in Figure~\ref{fig:limits}, existing instruments such as NIRC2 and NaCo can only achieve an absolute magnitude limit of 12 at extremely large separations.  For the separations used in this simulations, they have a contrast limit of $\sim 10^{-3}$ which will result in a low planet yield.

METIS is the best instrument for wide separations, quickly achieving a magnitude limit of 14 before which increases to 16 further out.  At very small separations, interferometric techniques such as VIKiNG are best.  Using our sample of simulated planets, we calculated the number of detectable planets at a given age and, from this, the detection probability.  The projected separation from Figure~\ref{fig:limits} is converted to physical separation by first multiplying by our distance of 140 pc.  This is then multiplied by 4/$\pi$ to take into account the average decrease in apparent separation due to projection effects.
\subsubsection{Entropy = 9.5\,$k_{B}$/baryon}
\begin{figure*}
\centering
\includegraphics[width=18cm]{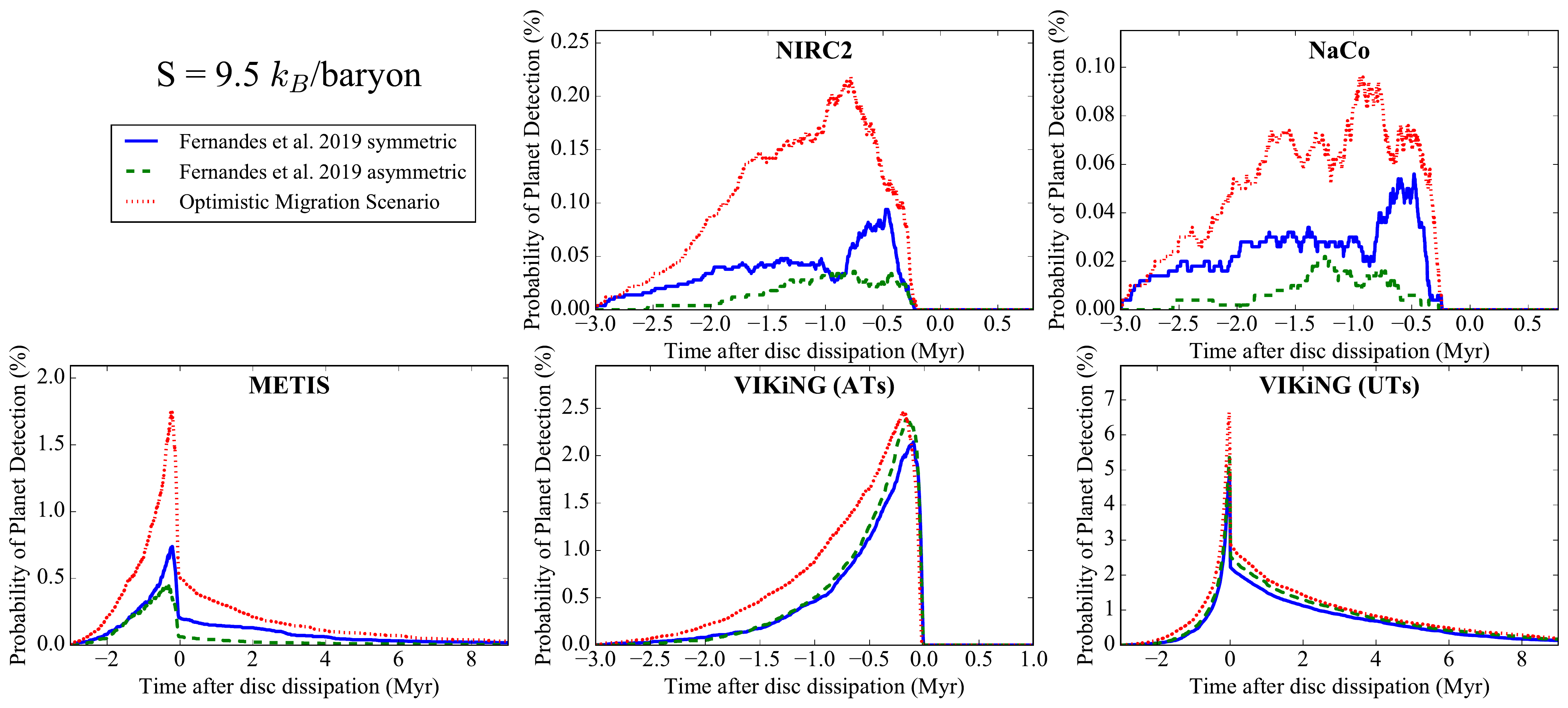}
\caption{Planet detection probability in the L' band for different instruments and techniques assuming a target of magnitude 8.3 at 140 pc, planet entropy of 9.5\,$k_{B}$/baryon and the distributions shown in Figures~\ref{fig:dist} and~\ref{fig:distMigration}.  NIRC2 and NaCo can barely see any planets still in the process of forming while METIS can see planets shortly after formation.  VIKiNG with the unit telescopes shows the greatest probability for both distributions and can see planets long after they form.}
\label{fig:realProbFernandesL9.5}
\end{figure*}

Figure~\ref{fig:realProbFernandesL9.5} shows the probability of planet detection with each instrument for different system ages using the distributions shown in Figures~\ref{fig:dist} and~\ref{fig:distMigration} and an internal entropy of 9.5\,$k_{B}$/baryon.

As shown in Figure~\ref{fig:realProbFernandesL9.5}, NIRC2 and NaCo are only able to detect some of the most massive planets during the runaway accretion phase.  VIKiNG and METIS are expected to produce a higher yield than both of these current instruments but since VIKiNG is more sensitive at small separations, this shows the most promise for future detections, regardless of planet distribution.  The probability peaks at the time the disc dissipates since this is the time all planets have formed and are at their brightest.  This is simply due to the fact that all of our systems have the same gas accretion time (3 million years.)  If instead we used a distribution of accretion times, the probability curves would not have this peak.
\subsubsection{Entropy = 10.5\,$k_{B}$/baryon}
\begin{figure*}
\centering
\includegraphics[width=18cm]{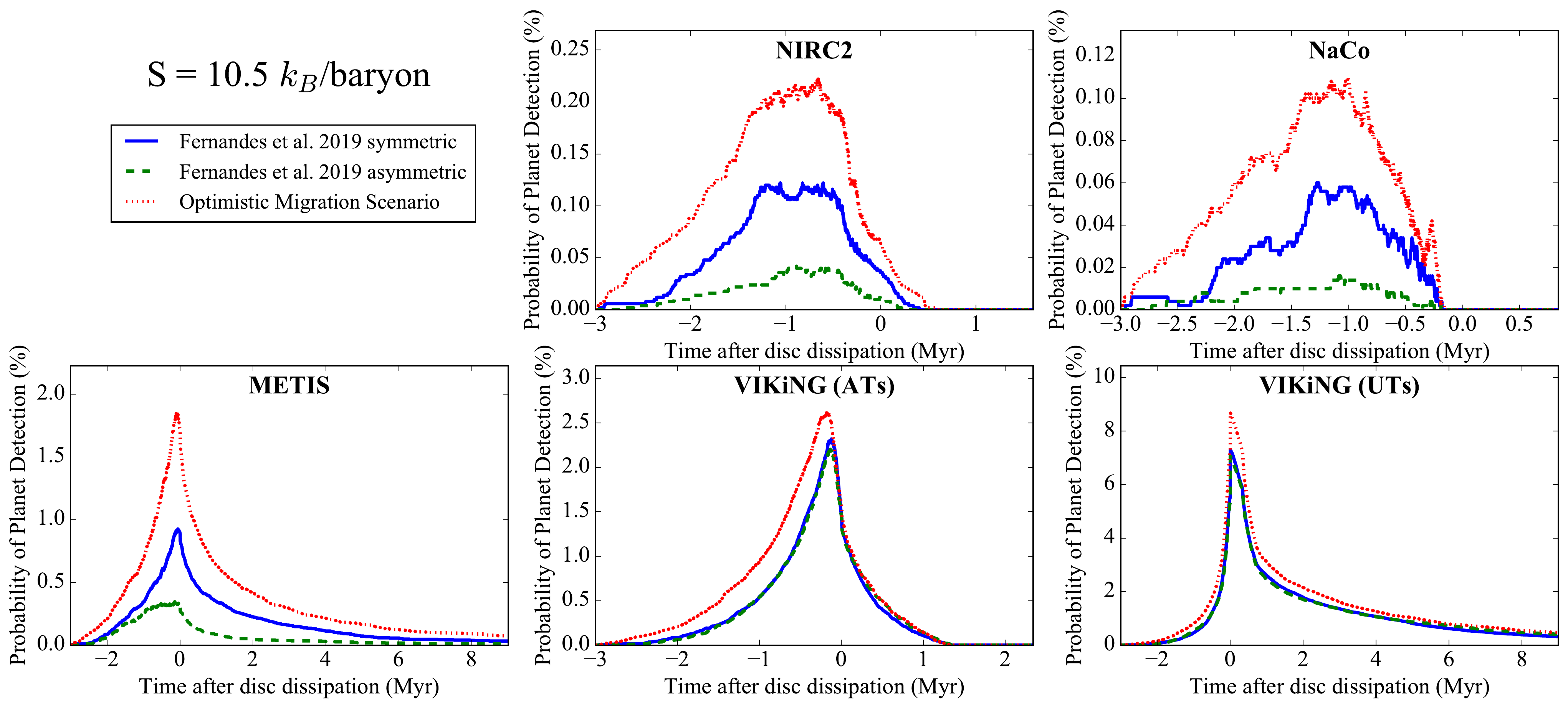}
\caption{Planet detection probability in the L' band for different instruments and techniques assuming a target of magnitude 8.3 at 140 pc, planet entropy of 10.5\,$k_{B}$/baryon and the distributions shown in Figures~\ref{fig:dist} and~\ref{fig:distMigration}.  With these hotter start planets, NIRC2 and VIKiNG with the ATs can now see some of them after formation.  As expected, detection probabilities have improved for all instruments.}
\label{fig:realProbFernandesL10.5}
\end{figure*}

This calculation was repeated, this time setting all planets' internal entropies to be 10.5\,$k_{B}$/baryon.  Figure~\ref{fig:realProbFernandesL10.5} shows the resultant detection probabilities.

As shown in Figure~\ref{fig:realProbFernandesL10.5}, increasing the entropy greatly increase planet brightness and detectability.  NIRC2 can see some of these planets shortly after formation, as can VIKiNG with the auxiliary telescopes.  The detection probabilities for METIS have also improved but VIKiNG with the unit telescopes still shows the greatest potential.
\subsection{K-band}
\begin{figure}
\centering
\includegraphics[width=8cm]{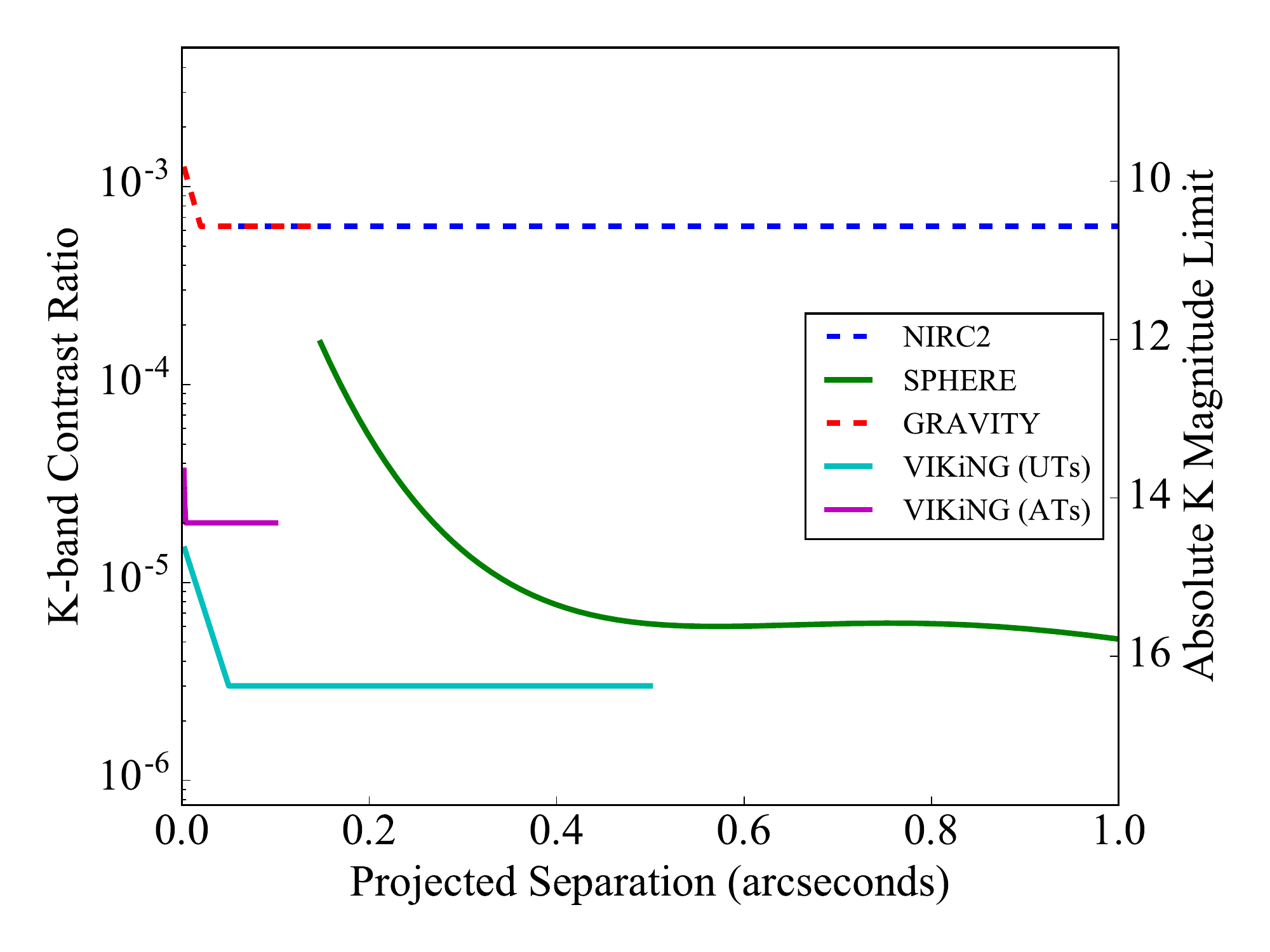}
\caption{K band contrast and magnitude limits for example instruments assuming a target of magnitude 8.3 at 140 pc.}
\label{fig:limitsK}
\end{figure}

Figure~\ref{fig:limitsK} shows the contrast and magnitude limits of NIRC2, SPHERE and GRAVITY in K-band assuming a target star of magnitude 8.3 at a distance of 140 pc.  The SPHERE limit was found using the IRDIS exposure time calculator using an input exposure time of 1 hour.  The NIRC2 and GRAVITY limits have not yet been demonstrated in K-band but are consistent with demonstrated closure-phase based detection limits \citep{hinkley2011observational}.
\subsubsection{Entropy = 9.5\,$k_{B}$/baryon}
\begin{figure*}
\centering
\includegraphics[width=18cm]{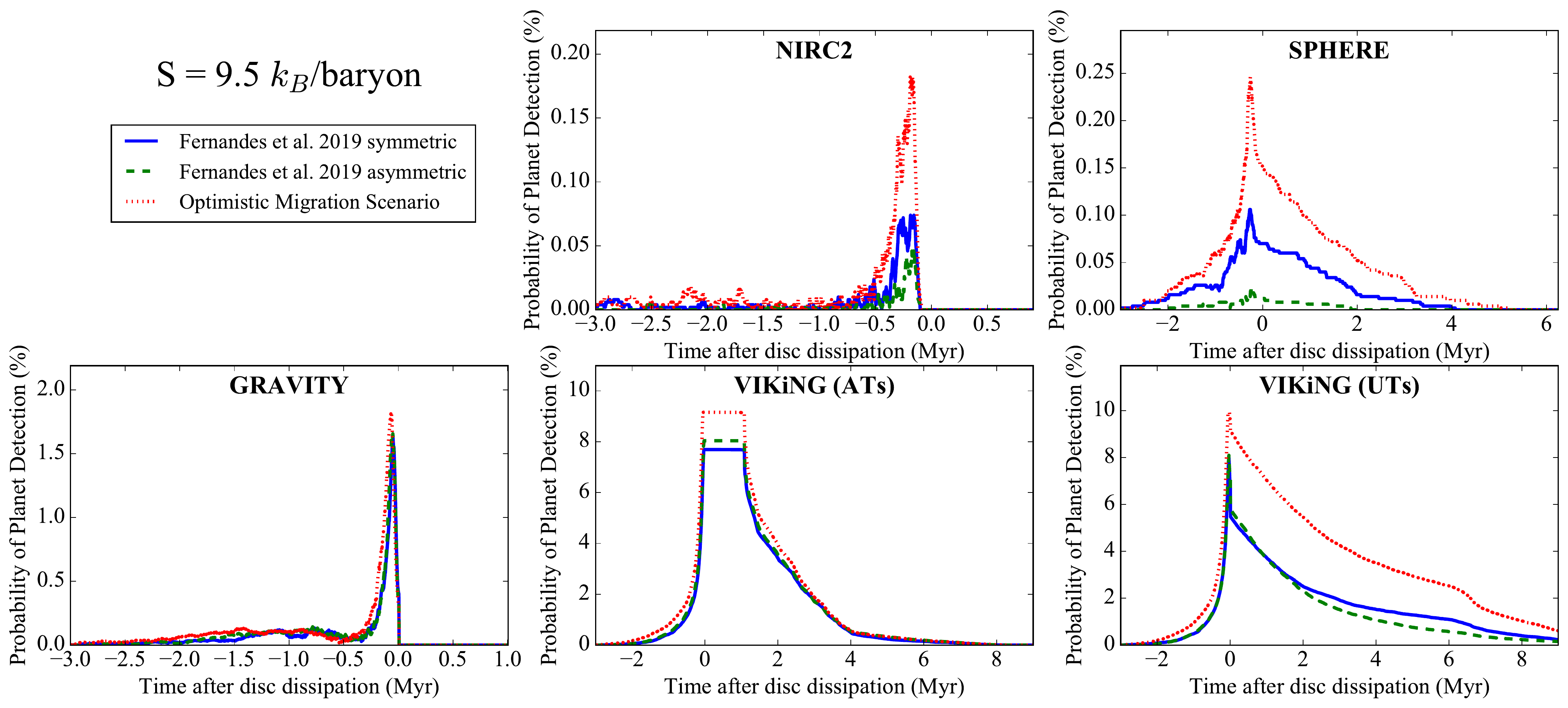}
\caption{Planet detection probability in the K band assuming a target of magnitude 8.3 at 140 pc, planet entropy of 9.5\,$k_{B}$/baryon and the broken power law distributions shown in Figures~\ref{fig:dist} and~\ref{fig:distMigration}.  For current instruments, SPHERE shows the greatest probability when assuming the optimistic migration power law while GRAVITY dominates for the asymmetric power law from \citet{fernandes2019hints}.  However, VIKiNG is predicted to be superior and is able to detect all planets in it's operational range just before and for $\sim$1\,Myr after formation.}
\label{fig:realProbFernandesK9.5}
\end{figure*}

Figure~\ref{fig:realProbFernandesK9.5} shows the detection probability in the K-band for NIRC2, SPHERE, GRAVITY and VIKiNG assuming a planet entropy of 9.5\,$k_{B}$/baryon.

The result in Figure~\ref{fig:realProbFernandesK9.5} shows, for current instruments, GRAVITY and SPHERE have the highest detection probability in the K-band, significantly surpassing NIRC2.  VIKiNG is again predicted to be superior to all current instruments and is able to detect all planets within its operational range (hence the flat top on the curve) up to $\sim$ one million years after formation.
\subsubsection{Entropy = 10.5\,$k_{B}$/baryon}
\begin{figure*}
\centering
\includegraphics[width=18cm]{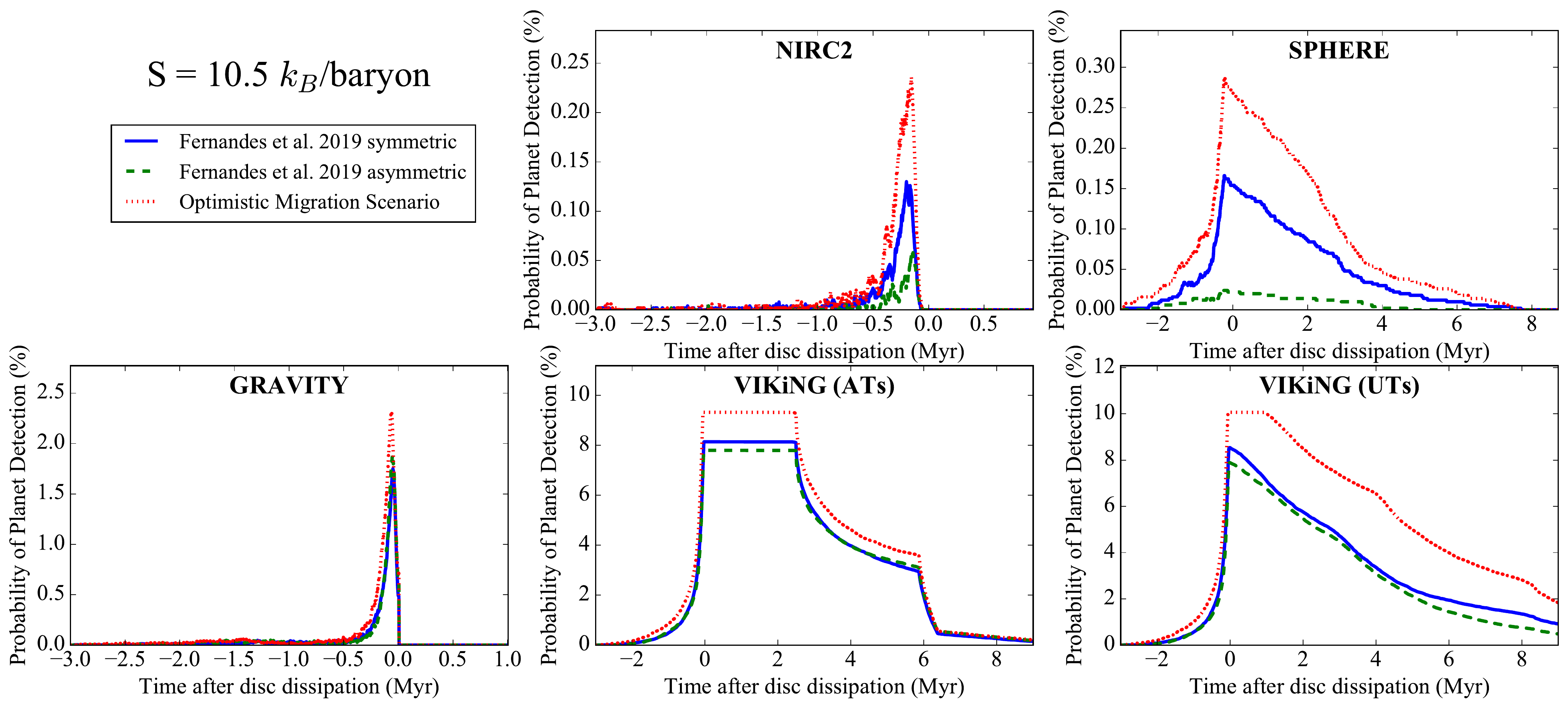}
\caption{Planet detection probability in the K band assuming a target of magnitude 8.3 at 140 pc, planet entropy of 10.5\,$k_{B}$/baryon and the broken power law distributions shown in Figures~\ref{fig:dist} and~\ref{fig:distMigration}.  Similar to L' band, increasing the entropy increases the detection probabilities for all instruments and VIKiNG with the UTs maintains a steady probability for a range of planet ages.}
\label{fig:realProbFernandesK10.5}
\end{figure*}

This is again repeated for internal entropies of 10.5\,$k_{B}$/baryon as shown in Figure~\ref{fig:realProbFernandesK10.5}.

Figure~\ref{fig:realProbFernandesK10.5} demonstrates once again that higher planet entropy results in higher detection probability.  VIKiNG with the UTs and ATs will be able to detect planets several million years after formation.  The probability for the Ats quickly drops off after two million years due to the high number of low mass planets fading below the detection limit.

In this paper we do not consider M-band because, although we expect the planets to be brighter at these longer wavelengths \citep{spiegel2012spectral}, we are working with solar-type stars which are too faint compared to the bright background in M-band to produce competitive contrast limits.
\section{Conclusions}
In this paper, we have linked known disc lifetimes and distributions of giant planets/disc radii to formation time, luminosity and fluxes of giant planets as they evolve.  From this, we can conclude that giant planets are unlikely to be detectable in the near infrared until rapid accretion when the planet resides in a gap.

Given our input parameters, it is no surprise that existing surveys have failed to find large numbers of giant planets consistent with core-accretion models.  However, we show that as long as we can get contrasts of more than 8 magnitudes in L'-band for a typical star in Taurus, instruments such as VIKiNG and METIS should be able to detect planets at a rate of greater than 1\%.  In the K-band, VIKiNG has an even greater chance of detecting planets after formation.  In all cases, hot-start planets have a greater detection probability.  Given a large enough sample (>100) of newly formed stars, we expect future instrument concepts such as METIS and VIKiNG to detect young planets during and after formation.

A limitation of this work so far is that it only considers solar mass stars and puts them all at the same age, distance and magnitude.  Future work should extend this simulation to a realistic sample of stars (such as the Taurus Molecular Cloud or Sco-Cen association) which will include a range of stellar mass, age and magnitude as well as protoplanetary disc lifetime, mass and size. Additionally, the planet distribution is expected to be further constrained over the coming years with a combination of radial velocity linear trends and direct imaging follow-ups to eliminate wide stellar companions, and also using Gaia astrometry. With these anticipated new data sets it will be possible to revisit the expected core-accretion giant planet direct detection survey yields. 

\section*{Acknowledgments}
We are grateful to Kaitlin Kratter for useful discussions on planetary interiors and entropy which helped improve these models, as well as feedback from Adam Kraus and Sarah Maddison on the manuscript. This work was supported by the Australian Government through the Australian Research Council's Discovery Projects funding scheme (project DP17010223).
\renewcommand\refname{References}
\bibliographystyle{mnras}
\bibliography{references}


\bsp	
\label{lastpage}
\end{document}